\documentclass[]{aa} 
\begin{document}
\def\rchi{{${\chi}_{\nu}^{2}$}}
\newcommand{\pcmsq} {cm$^{-2}$}
\newcommand{\xmm} {\sl XMM-Newton}
%
\title{XMM-Newton observations of OY Car III: OM light curve modelling, X-ray timing and spectral studies}
\titlerunning{XMM-Newton observations of OY Car III}
\authorrunning{Hakala \& Ramsay} 


\author{Pasi Hakala\inst{1} and Gavin Ramsay \inst{2}}

\offprints{P. Hakala}

\institute{$^{1}$Observatory, P.O. Box 14, FIN-00014 University of Helsinki,
Finland.\\
$^{2}$Mullard Space Science Laboratory, University College London,
Holmbury St Mary, Dorking, Surrey, RH5 6NT, UK.}

\date{}

\abstract{ We revisit the {\sl XMM-Newton} observations of the dwarf
nova OY Car taken in July 2000 which occured shortly after an
outburst. Ramsay {\it et al} (2001a) found a prominent energy
dependent modulation at a period of 2240 sec: this modulation was only
seen for $\sim$1/3 of the observation duration. In our new analysis,
we examine this time interval in greater detail. In addition to the
2240 sec period we find evidence for other periods, the most prominent
being near 3500 sec. Both these modulations are most likely due to
changes in photoelectric absorption over this period: this is
supported by phase-resolved spectroscopy. This may indicate the
presence of matter above the accretion disc or a presence of a
magnetic accretion curtain. In this case the 2240 sec period could
represent a spin period of the white dwarf and the 3500 sec period a
beat period between the spin and orbital periods. We also model the B
band and UV eclipse profiles and light curves using a new technique to
map the spatial extent of the accretion disc. As a result we find that
whilst the optical emission is dominated by both the emission close to
the accretion disc boundary layer and the hot spot where the accretion
stream hits the disc, the UV emission is mainly dominated by the inner
disc/boundary layer only.  \keywords{accretion, accretion discs --
binaries: eclipsing-- stars: individual: OY Car -- novae, cataclysmic
variables -- X-rays: stars} }

\maketitle
\section{Introduction}

Dwarf novae are interacting stellar binary systems in which material
gets transfered from a red dwarf secondary star onto a white dwarf via
Roche lobe overflow. In the absence of a strong magnetic field this
material forms an accretion disc around the white dwarf. Dwarf novae
show outbursts which occur on week to month timescales. For systems
which are viewed close to the plane of the binary system, eclipses of
the white dwarf by the secondary star are seen every orbital period.
These systems are important since they allow a detailed study of the
accretion disc as the secondary gradually covers then uncovers the
disc.

One such eclipsing dwarf nova is OY Car. It has a binary orbital
period of 1.51 hrs and has been extensively studied at various
wavelengths. Recently, it has been observed at X-ray and optical/UV
energies using {\sl XMM-Newton}. Ramsay et al (2001a) found that there
was a quasi-stable modulation of the X-rays at $\sim$2240 sec, which
was most prominent at lower energies. They speculated that this was
the due to disc material orbiting above the binary orbital plane or
due to the spin period of the white dwarf. In the latter scenario,
this would represent the first time that the spin period of the white
dwarf in a dwarf nova had been detected.

The eclipse profile shows a sharp drop in X-rays and in the UV
band. In X-rays the flux remains at a very low, but significant level
after this rapid (20--30 sec) drop. In the UV and B bands, the flux
decreases at a more gradual rate after the rapid drop since the
accretion disc is still visible after the eclipse of the white dwarf.
Wheatley \& West (2003) further analysed the {\xmm} data and found
that the X-ray eclipse lasted 30$\pm$3sec and concluded the X-ray
emission originated at the poles of the white dwarf.

In contrast, in this paper we investigate the nature of the 2240 sec
modulation in more detail by revisiting the {\sl XMM-Newton} data. In
particular we have carried out phased resolved spectroscopy over this
modulation period. Further, we apply a new eclipse mapping technique
to model the spatial distribution of the accretion disc using the OM
{\xmm} data.

\section{Observations}

{\sl XMM-Newton} was launched in Dec 1999 and has the largest
effective area of any imaging X-ray telescope. It has 3 medium
spectral resolution CCD type cameras on-board: two EPIC MOS detectors
and one EPIC pn detector. Since the EPIC pn detector has the higher
sensitivity we only discuss data taken using this detector
(Str\"{u}der et al 2001). In addition, it has an 0.3m optical/UV
imaging telescope which enables simultaneous X-ray optical/UV
observations (Mason et al 2001).

OY Car was observed during the performance verification phase of the
mission on June 29-30th 2000. The details of the observation are
reported in Ramsay et al (2001a,b). However, the main points to note
are that these observations were made some 4 days after an optical
outburst. The system had already faded by $V\sim$4 mag from its peak,
but it was still relatively bright in X-rays ($\sim$1.1 ct/s in the
EPIC pn). The exposure in the EPIC pn detector was 48ksec, it was
configured in full frame mode and the medium filter was used. The
optical data were obtained in the UVW1 filter (effective wavelength
2910 \AA). The data were processed using the {\sl XMM-Newton} Science
Analysis System v5.3.3. Additional optical data in B band was obtained
in August 2000 when it was at a similar brightness to the June
observation (Ramsay et al 2001a).

\section{X-ray Light Curve Analysis}

\begin{figure}
\begin{center}
\setlength{\unitlength}{1cm}
\begin{picture}(8,9)
\put(-0.5,-0.5){\includegraphics{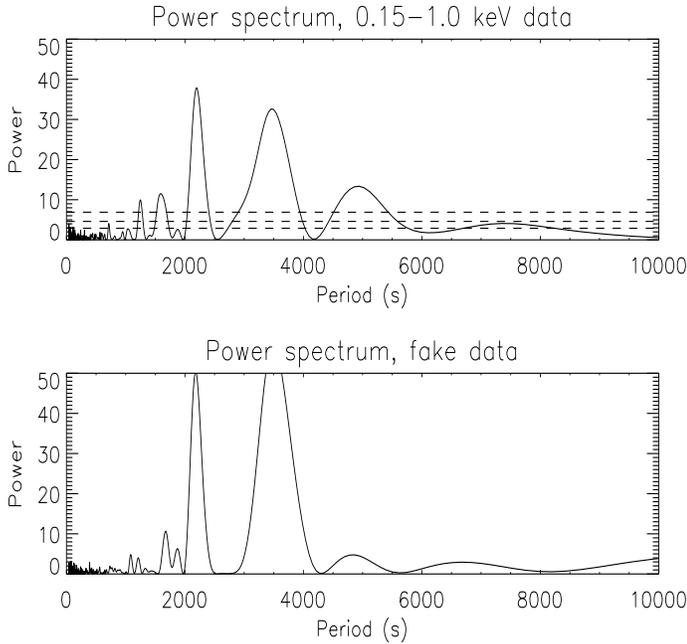}}
\end{picture}
\end{center}
\caption{The 0.15-1.0 keV power spectrum with 95\%, 99\% and 99.9\%
confidence limits (three dashed lines) from our simulations (top) and
a power spectrum produced using simulated fake data that contains the
two strongest signals (2193 sec and 3510 sec) seen in the power
spectrum. The shapes of 2193 sec and 3510 sec pulses are estimated
using a simultaneous 2nd order fourier fit with two periods.}
\label{pow-signi}
\end{figure}


\begin{figure}
\begin{center}
\setlength{\unitlength}{1cm}
\begin{picture}(8,10)
\put(-0.5,-0.5){\includegraphics{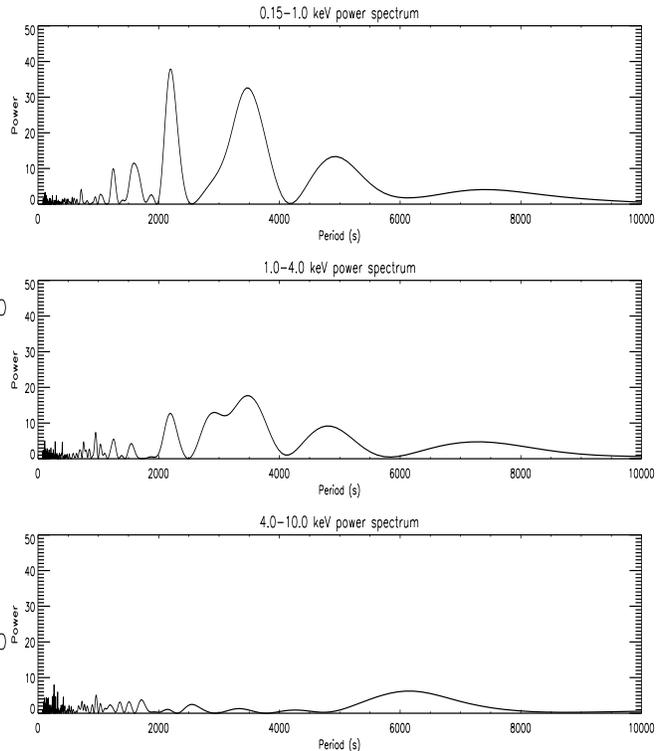}}
\end{picture}
\end{center}
\caption{The  power spectra at 0.15-1.0 keV (top), 1.0-4.0 keV (middle) and
4.0-10.0 keV (bottom).}
\label{pow-energy}
\end{figure}

\begin{figure}
\begin{center}
\setlength{\unitlength}{1cm}
\begin{picture}(8,12)
\put(-0.5,-0.5){\includegraphics{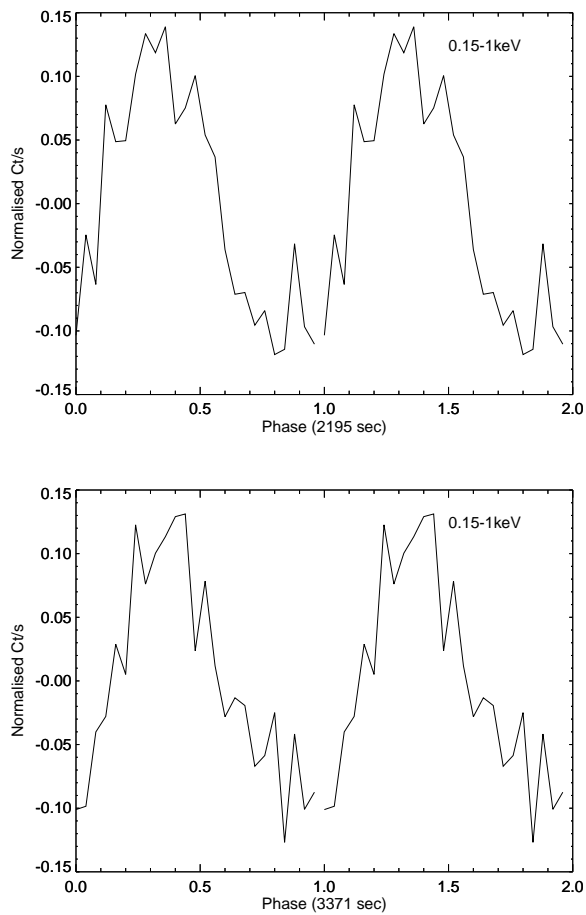}}
\end{picture}
\end{center}
\caption{The 0.15-1.0 keV light curve prewhitened on the 3510 sec period and
then folded on the 2193 sec period (top panel), and vice-versa
(bottom panel).}
\label{folds}
\end{figure}

Ramsay et al (2001a) found evidence for a quasi-stable X-ray
modulation on a period of $\sim$2240 sec, which was strongest in soft
X-rays. This modulation was observed for around 1/3 of the duration of
the observation and it was speculated that this period was related to
the spin period of the white dwarf.

We have re-examined the light curves of OY Car: in particular we
perform a timing analysis of that portion of the light curve in which
the 2240 sec modulation is strongest (HJD 2451725.6548--2451725.8622)
in 4 energy bands (0.15--10.0, 0.15--1.0, 1--4 and 4--10keV). (We
removed the time of the eclipse from the light curves). We show the
amplitude spectrum of the 0.15--1.0keV light curve in Figure
\ref{pow-signi} (top panel). The most prominent peak has a period of
2193 sec: within the errors, this period is consistent with the 2240
sec period found by Ramsay et al (2001a). However, there is a second
prominent peak at 3510 sec: this peak is still present if we
pre-whiten the data on the 2193 sec period. We performed the same
analysis in the 2 more restricted energy bands (Figure
\ref{pow-energy}): we find that both the 2193 sec and 3510 sec periods
are more prominent at softer energies. A peak was seen at 3510 sec in
the amplitude spectrum shown in Ramsay et al (2001a), however, it is
more prominent in this new study since we have restricted the analysis
to the time interval in which the quasi-stable modulation is seen. In
order to demonstrate the prominence of both of these periods, we show
the data folded on either of the two periods, after they have been
prewhitened using the other period (Figure \ref{folds}).

To determine the significance of these two periods we randomly
populated the time bins with the observed count rates. We then
generated 10000 power density spectra: the 95\%, 99\% and 99.9\%
confidence limits from these are shown together with the 0.1-1.0 keV
power spectrum in Figure \ref{pow-signi}(top panel). It is clear that
both the 2193 and 3510 sec periods are highly significant. To
determine the errors on these periods we used the Cash statistic (Cash
1979), which implies: 2193$^{+17}_{-35}$ and 3510$^{+25}_{-104}$
sec. There are three other periods which are significant at $>99.9$\%
level: 1248$^{+17}_{-15}$ sec, 1574$^{+18}_{-21}$ sec and
4844$^{+120}_{-104}$ sec.

It is possible that some of these less significant periods could be
produced as a result of aliasing between the most prominent periods
(2193 sec and 3510 sec) and the window function or as a result of
these two periods having a non-sinusoidal pulse shape. In order to
investigate this further we have simulated a dataset with the same
time points as the original data. However, this fake data only
contains the two strongest signals (2193 sec and 3510 sec). The shape
of these signals is obtained from the real data by using a
simultaneous two-period fourier fit, where each of the two pulse
shapes are represented with a second order fourier series. As a result
we find that at least the 1574s period could be produced by the
effects mentioned above, Figure \ref{pow-signi} (lower panel).

We discuss the possible origin of all of these periods in \S \ref{mod_discuss}.

\section{X-ray Spectral Analysis}
\label{spec}

Taking the same time interval that the quasi-stable behaviour is
present as in the previous section, we phased events on the 2193 and
3510 sec quasi-stable periods reported in the previous
section. Initially we extracted 6 spectra covering different 2193 sec
phase intervals. In extracting these spectra we used apertures
$\sim30^{'}$ in radius centered on OY Car, chosen so that the aperture
did not cover more than one CCD.  This encompasses $\sim$90 percent of
the integrated PSF (Aschenbach et al 2000). A background spectrum was
extracted from the same CCD on which the source was detected, scaled
and subtracted from the source spectra.

Since the response of the detectors is not well calibrated at present
below $\sim$0.2keV, energies below this were ignored in the following
analysis. The response file epn$\_$ff20$\_$sdY9$\_$medium.rmf was
used: this response includes single and double pixel events.

Ramsay et al (2001b) found that they could not well fit the integrated
EPIC spectra of OY Car with a single temperature thin plasma model. A
multi-temperature model gave a much better fit to the data. In their
analysis Ramsay et al (2001b) concluded that a best fit for the
integrated spectrum was obtained with a three temperature plasma model
with a partially covering absorber. The multi-temperature nature was
supported from line H/He-like intensity ratios of Iron and Sulphur and
also by the various line species seen in the high resolution RGS
spectrum.

\begin{figure}
\begin{center}
\setlength{\unitlength}{1cm}
\begin{picture}(8,12)
\put(-0.5,-1.5){\includegraphics{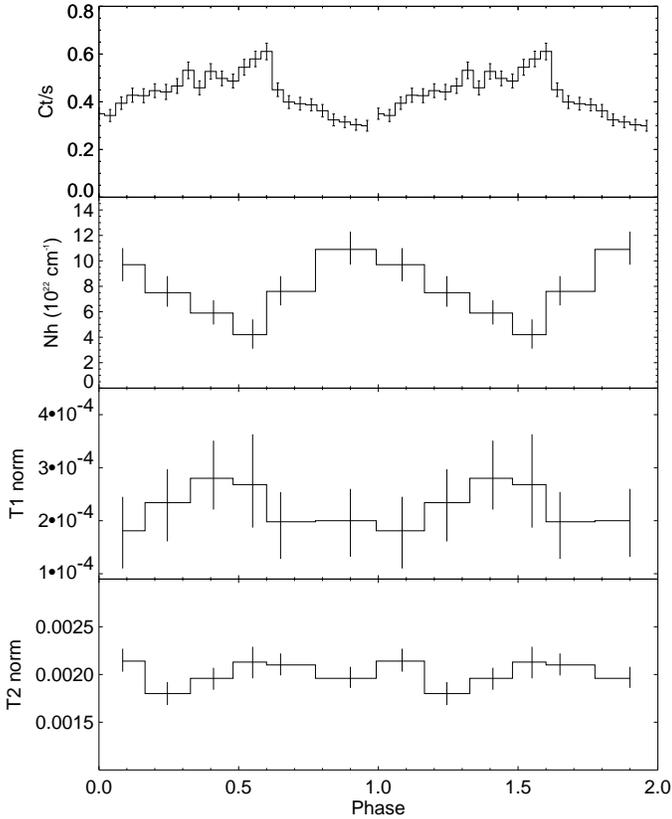}}
\end{picture}
\end{center}
\caption{From the top panel: using the time interval which showed
evidence for a modulation on the 2193 sec period, we fold that data on
that period in the 0.15-1.0keV energy band; the fitted neutral
absorption column; the normalisation of the lower temperature
component ($kT_{1}$); the normalisation of the higher temperature
component ($kT_{2}$).}
\label{fold}
\end{figure}

\begin{figure}
\begin{center}
\setlength{\unitlength}{1cm}
\begin{picture}(8,7.0)
\put(-0.5,-0.0){\includegraphics{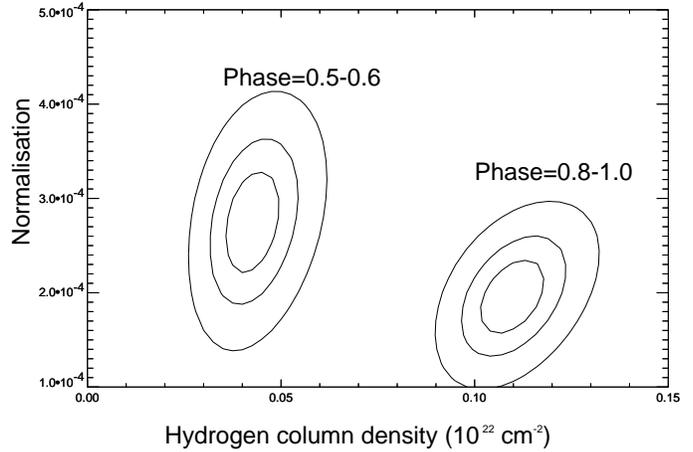}}
\end{picture}
\end{center}
\caption{Confidence contours in the ($N_{H}$, $T_{1}$,normalisation)
plane for the phase minimum and maximum spectra.
The contours correspond to 68\%, 90\% and 99\% limits.}
\label{cont}
\end{figure}

\begin{figure}
\begin{center}
\setlength{\unitlength}{1cm}
\begin{picture}(8,6)
\put(-0.5,-0.5){\includegraphics{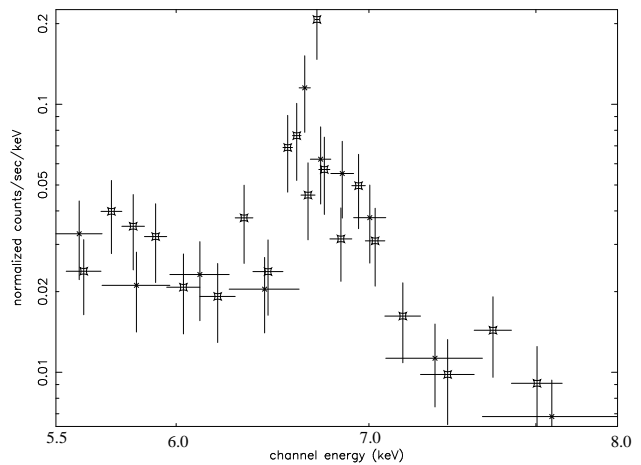}}
\end{picture}
\end{center}
\caption{The Fe K$\alpha$ complex at the 2193 sec phase minima
(0.5--0.6) (shown an asterix) and maxima (0.9--1.0) (shown as stars).}
\label{fespec}
\end{figure}

We fitted each of the 6 phase resolved spectra separately using the
X-ray fitting package {\tt XSPEC}. Each spectrum was binned so that a
minimum of 40 counts were present in each spectral bin. An initial
analysis showed that a two temperature fit gave much better fits to
the data. In these phase resolved spectra adding a third temperature
component did not significantly improve the fit. Further, we find that
the temperatures for each of thermal components were consistent in
each of the 6 phase resolved spectra (kT$_{1}$=1.4keV and
kT$_{2}$=10keV). We therefore fixed the temperatures of the thermal
plasma models at these temperatures. In each fit the free parameters
were the absorption (we used a simple neutral absorption model) and
the normalisation. The metal abundance was fixed at solar.

The resulting fits are shown in tabulated form in Table \ref{fits} and
in graphical form (along with the light curve folded on the 2193 sec
period) in Figure \ref{fold}. The fits are reasonably good in the
phase intervals 0.32--0.80, but in the remainder it is less so
(\rchi$\sim$1.5). In the case of the phase interval 0.0--0.17 the fit
was significantly improved when we fix the metal abundance at 1.5
solar (\rchi=1.35). It is clear from Figure \ref{fold} that there is
no significant variability in the normalisations of the thermal plasma
components. However, there is a variation in the column of the neutral
Hydrogen density: it is anti-correlated with the folded intensity
light in that it shows a maximum at intensity minimum. We show in
Figure \ref{cont} the confidence contour plot in the ($N_{H}$,
$T_{low,norm}$) plane for the spectra covering the phase maximum and
minimum: there is a clear separation in the contours.

\begin{table}
\begin{center}
\begin{tabular}{llrrr}
\hline
Phase & $N_{H}$ & Norm $T_{1}$ & Norm $T_{2}$ & \rchi \\
& ($10^{20}$  & ($\times10^{-4}$) & ($\times10^{-3}$) & (dof)\\
& \pcmsq) &              &              &  \\
\hline
0.0-0.17 & 9.7$\pm$1.3 & 1.81$^{+0.64}_{-0.71}$ & 
2.14$^{+0.13}_{-0.11}$ & 1.50 ((62)\\
0.17--0.32 & 7.5$^{+1.3}_{-1.1}$& 2.34$^{+0.63}_{-0.73}$ &
1.80$\pm0.12$ & 1.60 (52) \\
0.32-0.50 & 5.9$^{+1.0}_{-0.9}$ & 2.80$^{+0.71}_{-0.59}$ &
1.96$^{+0.11}_{-0.12}$ & 1.27 (72\\
0.50-0.60 & 4.2$^{+1.2}_{-1.1}$ & 2.68$^{+0.95}_{-0.81}$ &
2.13$^{+0.16}_{-0.17}$ & 0.71 (42)\\
0.60-0.80 & 7.6$^{+1.2}_{-1.1}$ & 1.98$^{+0.56}_{-0.70}$ &
2.10$^{+0.12}_{-0.11}$ & 1.18 (75)\\
0.80-1.00 & 10.9$^{+1.4}_{-1.2}$ & 2.00$^{+0.60}_{-0.68}$ &
1.96$^{+0.12}_{-0.10}$ & 1.47 (67)\\
\hline
\end{tabular}
\end{center}
\caption{The results of phase resolved spectral fits folded on the
2193 sec period. The temperature of the cooler ($T_{1}$) plasma was
fixed at 1.4keV, the hotter ($T_{2}$) at 10keV.}
\label{fits}
\end{table}

We show in Figure \ref{fespec} the X-ray spectra near 6--7keV for
spectra taken from the phase maximum and minimum: the profile of the
Fe K$\alpha$ line does not differ significantly. This further
strengthens the argument that the spectral changes over the 2193 sec
period are most likely caused by the change in the amount of
absorption.
          
We also folded the data on the 3510 sec quasi-stable modulation: we
find a similar result to that found for folding on the 2193
sec. Namely, the temperature and normalisation of the two thermal
plasma components do not vary over phase, but the absorption
component varies significantly. For brevity we do not show the results
in Table or Figure form.

\section{Modelling the B band and UV data}

It is clear from Ramsay et al. (2001a) that the B and UV light
curves exhibit clear modulation over the orbital period as well as
asymmetric eclipse profiles. In this section we will describe and
discuss the modelling of those data in detail.

Our basic assumption is that all of the B and UV modulation is
caused by the changing visibility of different parts of (at least)
partially optically {\sl and} physically thick accretion disc and the
eclipse of such disc by the secondary star.  In our modelling we have
taken the idea of `fireflies'' presented in Hakala, Cropper \& Ramsay
(2002) and developed it further to suit the modelling of optically
thick accretion discs.
	  
Hakala et al. (2002) use a swarm of emitting point sources (fireflies)
to model the shape of an accretion stream in eclipsing AM Herculis
systems.  In their approach the swarm is regularized to favour
realistic `banana' shapes for the accretion stream (without fixing
the exact location or shape of the stream).  They then used a variant
of a genetic algorithm to find the optimal accretion stream shape and
location.

If we want to apply a similar approach to locating the emitting
regions in a disc accreting system, like OY Car, we need to make
several modifications to the original code. In fact, the only
assumptions that remain the same are the idea of `fireflies' as a
numerical model for the emission and the optimisation algorithm. We
next discuss our model computation in detail.

\subsection{The model}
	
\begin{figure} 
\begin{center} 
\setlength{\unitlength}{1cm}
\begin{picture}(8,16)
\put(-2.0,-10.0){\includegraphics{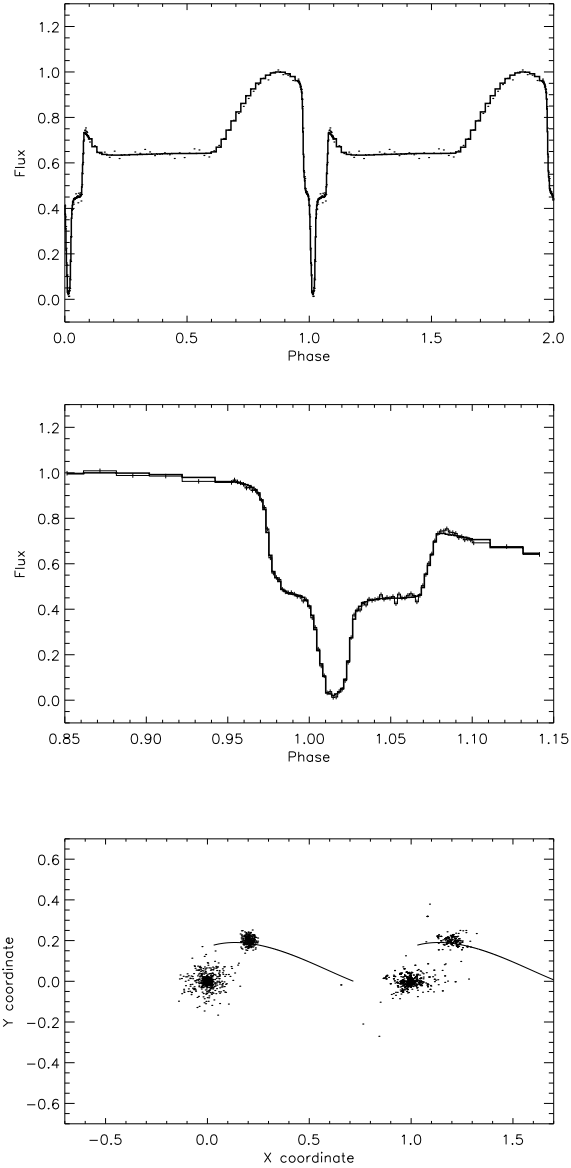}} 
\end{picture} 
\end{center} 
\caption{Fits to the
synthetic data. The whole light curve (top); zoomed in
around the eclipse (middle); and the original (bottom, left)
and reconstructed (bottom, right) brightness distributions
for the fit. The bottom panel shows the ballistic stream trajectory and
an accretion disc looking down on the orbital plane.}  
\label{synth}
\end{figure}

We first restrict the possible emission to originate within an
accretion disc like volume around the primary. The boundary conditions
for this volume are: maximum radius 0.45a, maximum height $\pm
0.02a$. In addition we define a somewhat thicker disc like volume of
maximum height $\pm 0.025a$. This volume is assumed to have a
homogenous optical thickness, that remains constant throughout the
modelling.  The reason for this is to provide estimates for the
absorption from the different parts of the disc and thus guarantee an
optically thick solution.

To compute the optical light curve for any given distribution of
fireflies inside this boundary, the algorithm for the light curve
generation procedes as follows. We first loop over all the orbital
phases to be fitted and for each phase:

\begin{figure}
\begin{center}
\setlength{\unitlength}{1cm}
\begin{picture}(8,16)
\put(-2.0,-10.0){\includegraphics{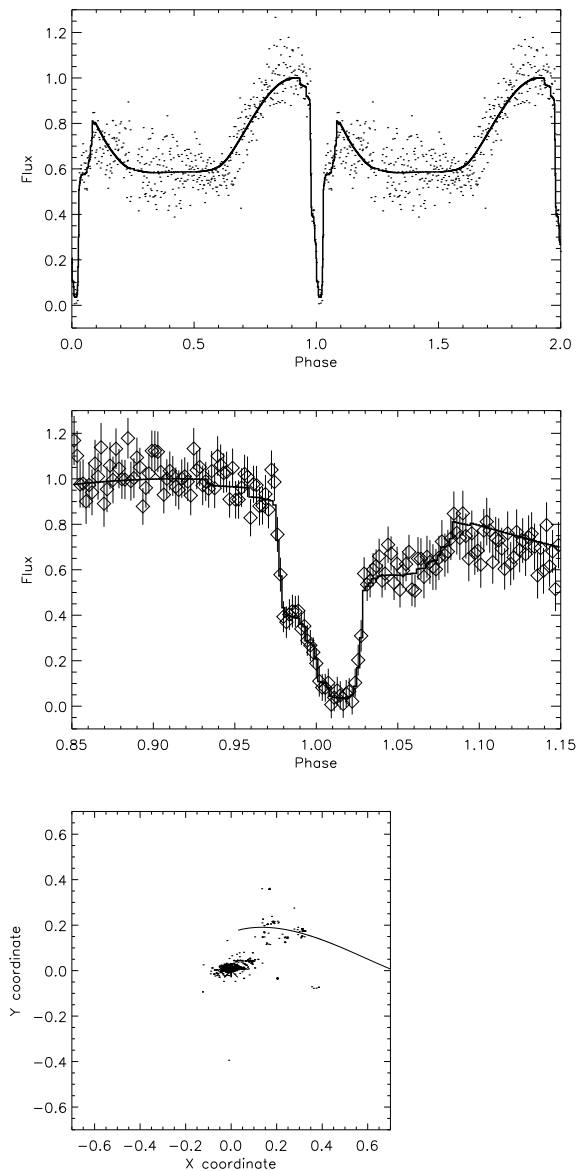}}
\end{picture}
\end{center}
\caption{Our best fit to the B-band data assuming {\sl no}
regularization. The whole light curve is shown on top and we have
zoomed in around the eclipse (middle). The resulting firefly
distribution together with a ballistic accretion stream trajectory is
plotted in bottom (Primary at the origin, L1 point at
approx. coordinates of [0.7,0.0]).}
\label{b-noreg}
\end{figure}

\begin{figure}
\begin{center}
\setlength{\unitlength}{1cm}
\begin{picture}(8,16)
\put(-2.0,-10.0){\includegraphics{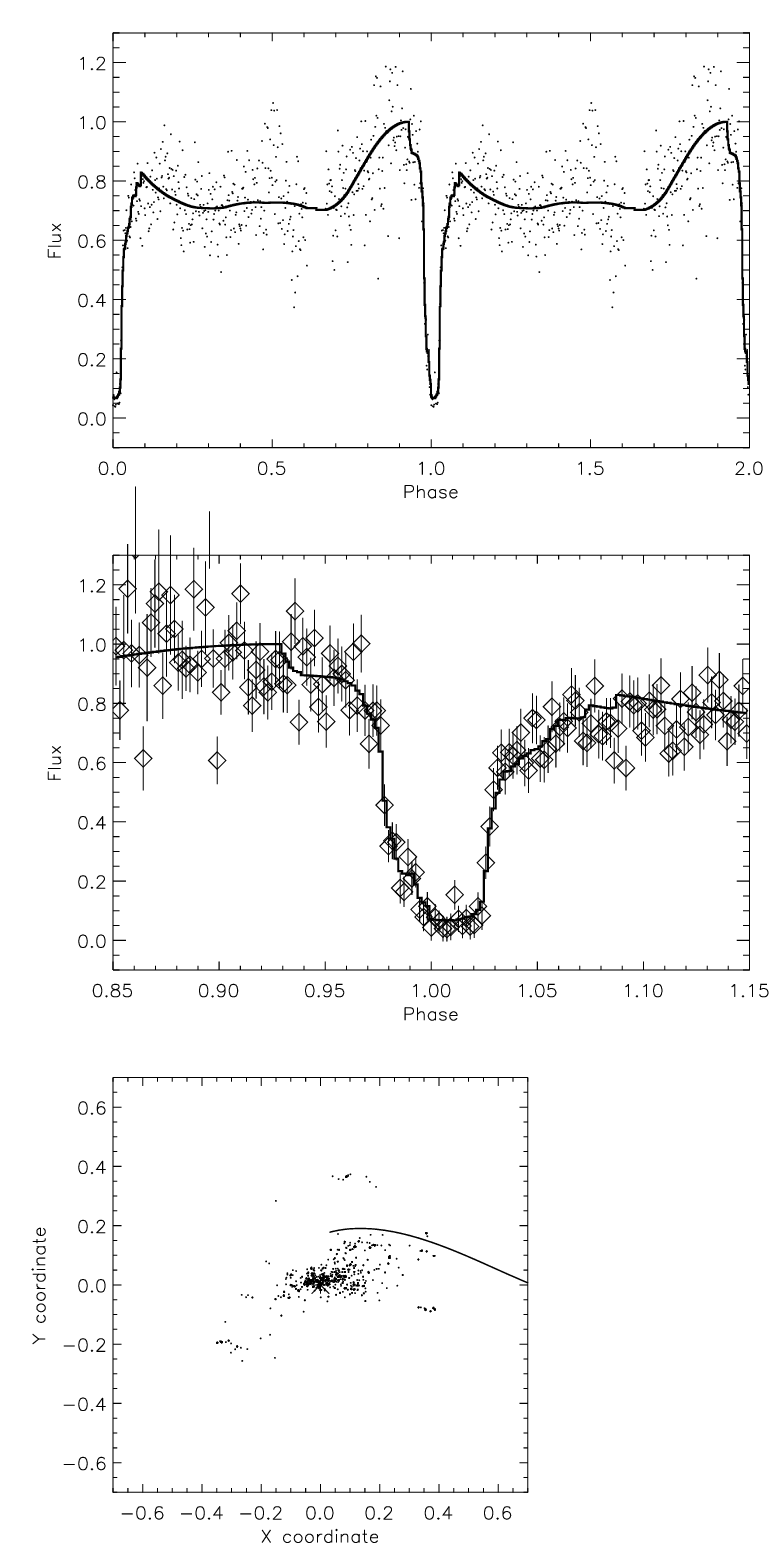}}
\end{picture}
\end{center}
\caption{Our best fit to the UV data assuming {\sl no}
regularization. The whole light curve is shown on top and we have
zoomed in around the eclipse (middle). The resulting firefly
distribution together with a ballistic accretion stream trajectory is
plotted in bottom (Primary at the origin, L1 point at
approx. coordinates of [0.7,0.0]).}
\label{uv-noreg}
\end{figure}

\begin{figure}
\begin{center}
\setlength{\unitlength}{1cm}
\begin{picture}(8,16)
\put(-2.0,-10.0){\includegraphics{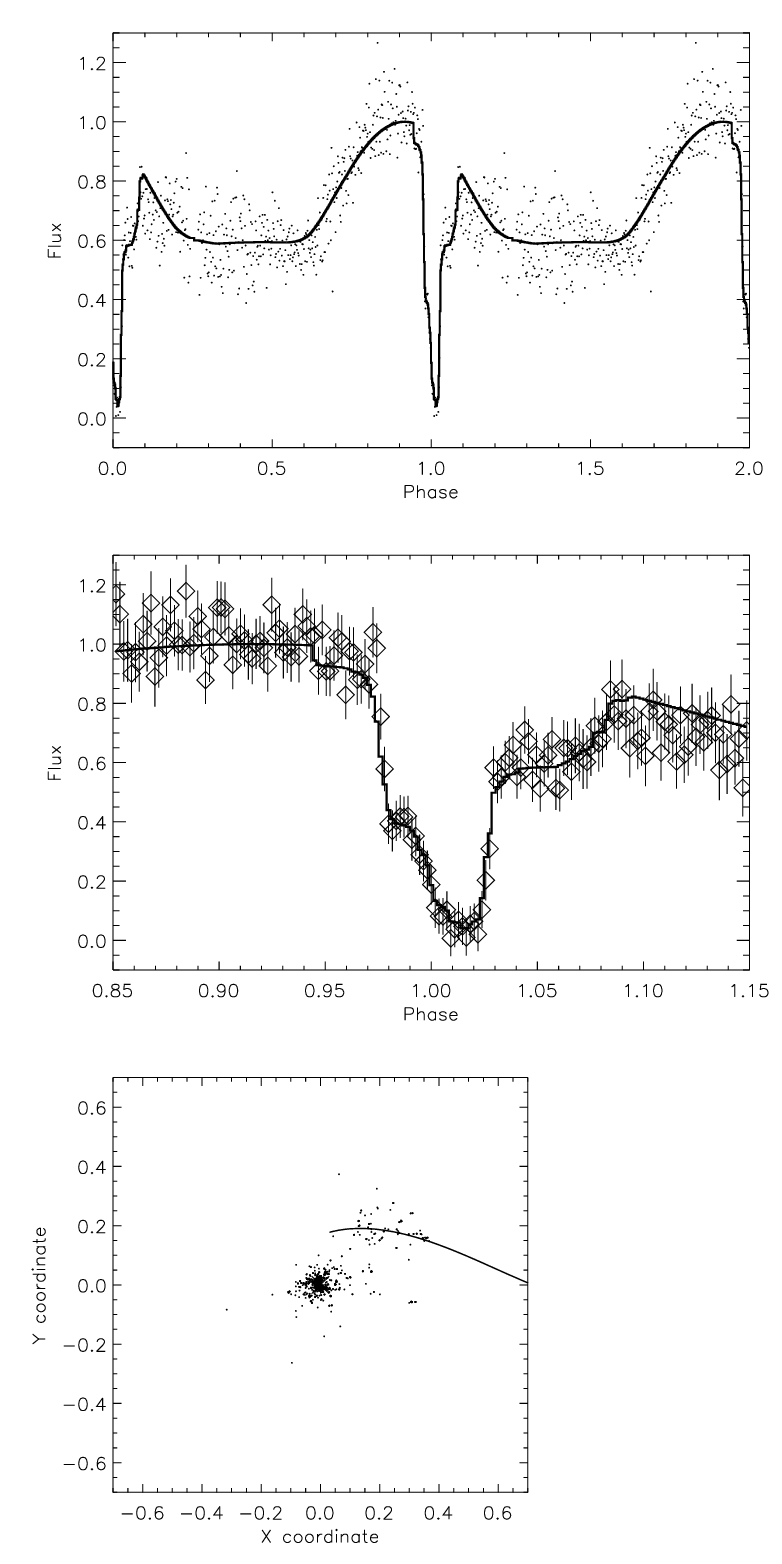}}
\end{picture}
\end{center}
\caption{Our best fit to the B-band data assuming axisymmetric
regularization. The whole light curve is shown on top
and we have zoomed in around the eclipse (middle). The resulting firefly
distribution together with a ballistic accretion stream trajectory  
is plotted in bottom (Primary at the origin, L1 point at approx. coordinates
of [0.7,0.0]).}
\label{breg}
\end{figure}

\begin{figure}
\begin{center}
\setlength{\unitlength}{1cm}
\begin{picture}(8,16)
\put(-2.0,-10.0){\includegraphics{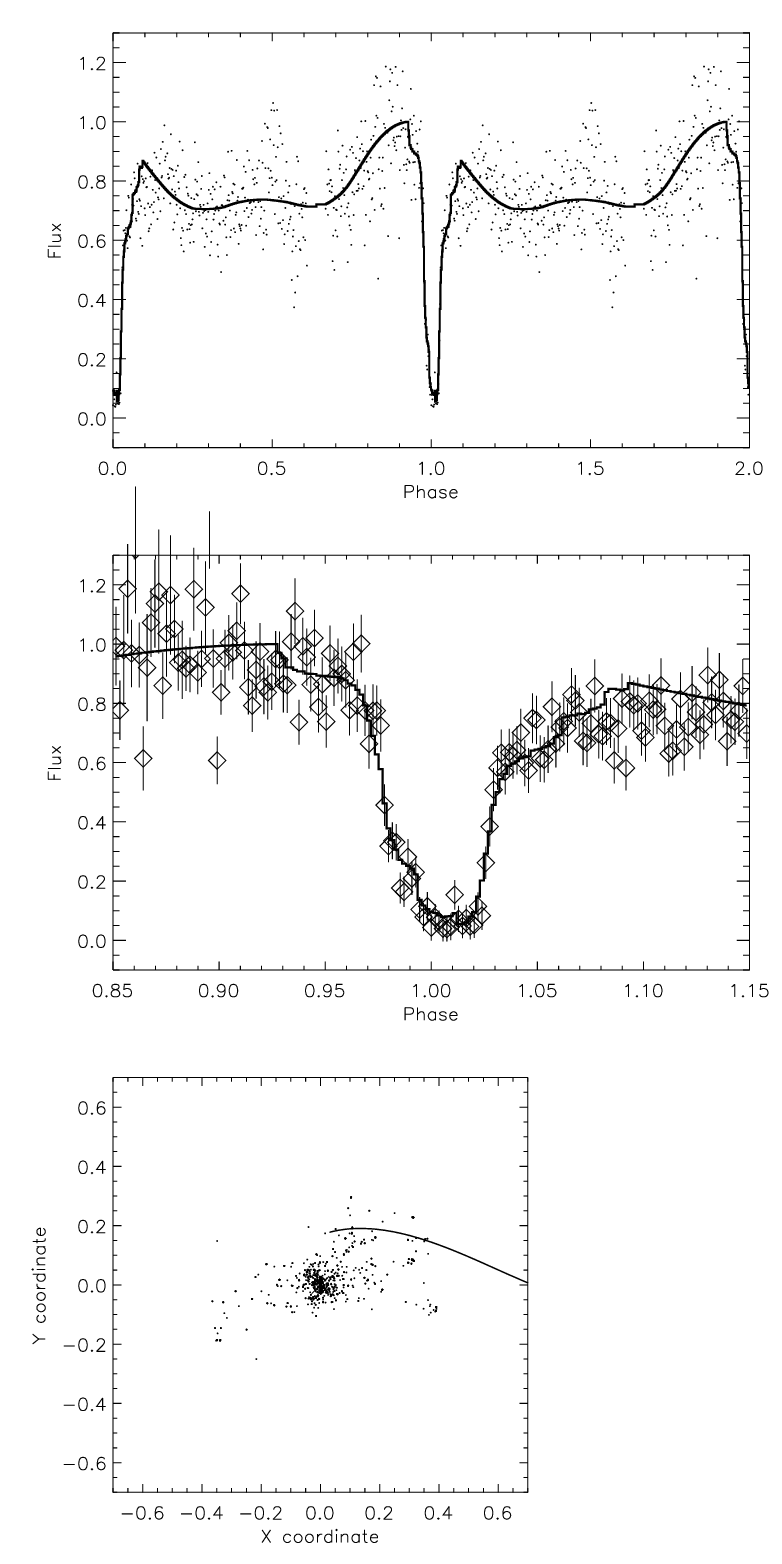}}
\end{picture}
\end{center}
\caption{Our best fit to the UV data (with axisymmetric
regularization). The whole light curve is shown on top
and we have zoomed in around the eclipse (middle). The resulting firefly
distribution
(bottom).}
\label{uvreg}
\end{figure}

\begin{itemize}

\item Select the fireflies that are not eclipsed by the secondary;

\item Determine the absorption coefficient for each of the
selected fireflies.  This is done by computing the length of
a path that light from any particular fly has to travel
inside a predefined disc volume;

\item Sum up the emission from all visible flies at this
phase (taking into account the variable absorption for
different flies).

\end{itemize}

\begin{figure}
\begin{center}
\setlength{\unitlength}{1cm}
\begin{picture}(8,6)
\put(-2.0,-7.5){\includegraphics{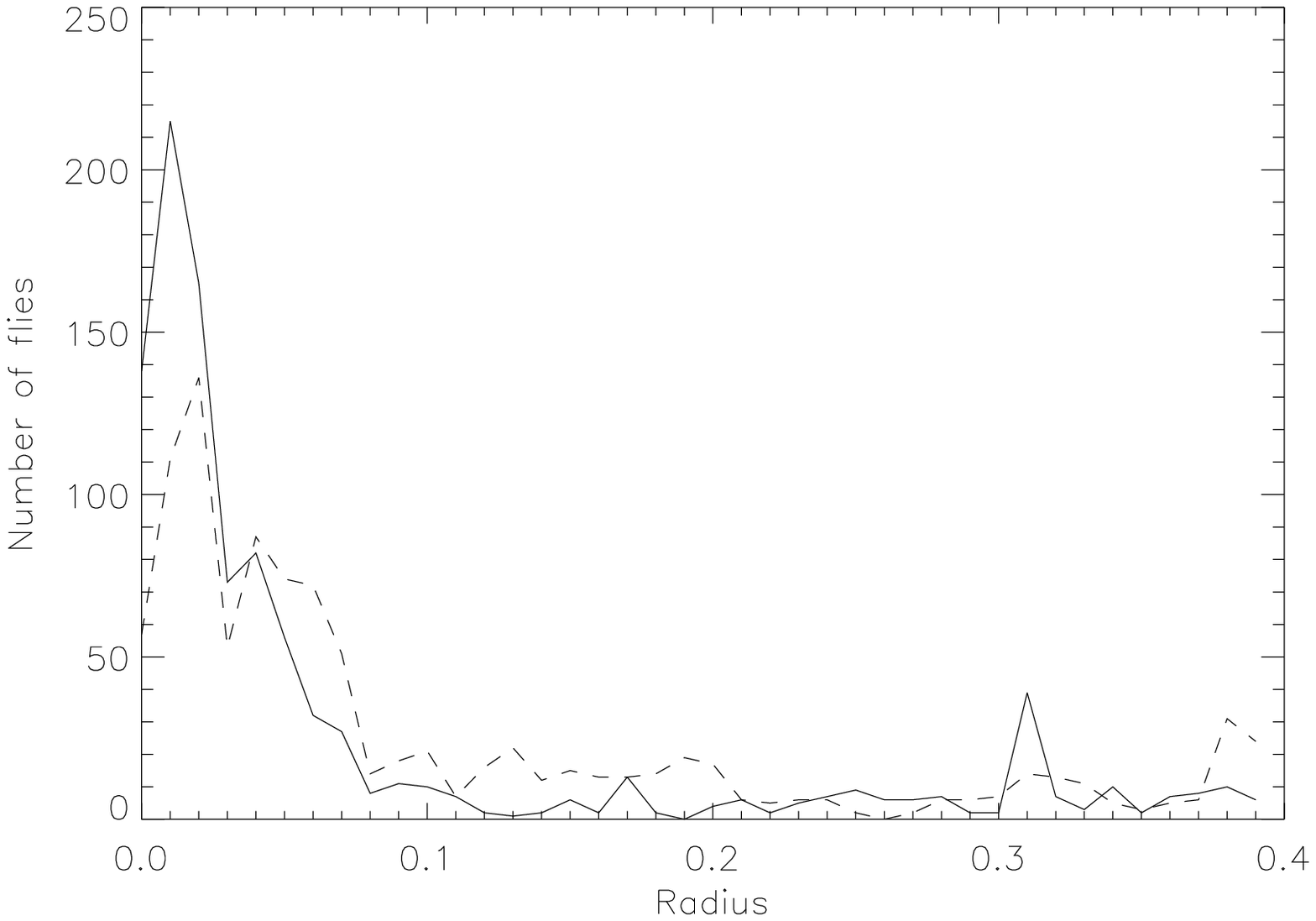}}
\end{picture}
\end{center}
\caption{The radial distribution of fireflies in B (solid) and UV  bands (dashed).
The hot spot in B band is visible at a distance of 0.3a.}
\label{fly-hist}
\end{figure}

The main difference between the `firefly approach' and the `normal
approach' is that the firefly approach is (in analogue with
hydrodynamics) a particle based solution whilst the `normal' approach
relies on fixed grid on which the solution is determined.  The reasons
behind our choice of the firefly approach here are the following ones.
Firstly, one of the main axioms of the standard eclipse mapping
routine (Horne 1985) is that the eclipse profile of the disc is solely
determined by the changing obscuration of the disc by the
secondary. Thus outside the eclipse phases the light curve is
constant.  This implies that the same algorithm could not be used to
fit the rest of the light curve.  Normally the out-of-eclipse
variation, often seen in quiescent state CV light curves, has to be
removed {\it before} the fitting, which can cause undesirable bias in
the results.  Rutten (1998) developed the eclipse mapping by including
a 3D shape for the disc.  His model could be used for out-of-eclipse
light curve modelling, but it has it's limitations as well, i.e. we
have to know and fix the 3D shape of the disc beforehand.

The advantage of the firefly modelling is the lack of any
grids. The flies are free to move within a predefined volume
and their final distribution represents the location or
distribution of emission within the disc. However, our
approach does require that the disc has an axisymmetric
homogenous density structure, as this is assumed for the
computation of optical thickness effects that dominate the
out-of-eclipse light curve shape.

\subsection{Model Regularization}

As in most of inversion problems, there are an infinite number of
solutions if no regularization is employed. 
In order to overcome this, we have (like many previous authors)
decided to search for the most axisymmetric image that can reproduce
the observations (axisymmetric about the accretion disc rotation
axis). In the case of the firefly approach this simply means a
regularization term, where we have defined the disc in $i$ sectors,
and counted the number of flies in each sector $N_{i}$ (here we have
chosen to use 36 sectors).  Then we maximize the entropy of such
division i.e. we first compute

\begin{eqnarray}
p_i  = \frac {N_i} { \sum_{i=1}^{36} N_i}
\end{eqnarray}

\noindent
where $p_i$is the fractional flux in each sector and the entropy term is

\begin{eqnarray}
S  =  \sum_{i=1}^{36} -p_i \log(p_i)
\end{eqnarray}

\noindent
finally our merit function to be minimized is

 \begin{eqnarray}
  F  =  \chi_\nu^2-\lambda S
   \end{eqnarray}   
   
    \noindent
     where $\chi_\nu^2$ is the standard reduced $\chi^2$ term, $\lambda$ is
the
Lagrangian multiplier
and $S$ the entropy term. 

\subsection{The fitting algorithm}

We carry out the actual fitting using the same genetic algorithm
variant as in Hakala et al. (2002). The only difference is that in
this case the initial population of solutions for fitting consists of
disc like swarms of random fireflies. These are then evolved in
similar fashion to fit the eclipse profile and the overall light curve
simultaneously. As the whole light curve has 532 points in B and 514
points in UV, we have decided to emphasize the eclipse phases
(0.95-1.05) by summing the Chi-square values from this phase range
three times. We typically use 50 individual swarms in each generation
and after around 1500 generations, the fits stop evolving in a
reasonable time.

In order to test our approach we have created a synthetic light curve
using a known disc brightness distribution, added appropriate levels
of noise and fitted that. The results of this exercise are shown in
Figure \ref{synth}. The top panel shows the whole orbital light curve
together with the fit, the middle panel the phase range close to the
eclipse phase and the bottom panel the synthetic brightness
distribution (left) and the reconstructed distribution (right). The
fits to the synthetic data are encouraging, with the reconstructed
disc being close to the input disc. There is, however, a small amount
of smearing, which is most likely caused by finite phase resolution,
regularization and/or noise in the data.

\subsection{The Results}

We have plotted in Figures \ref{b-noreg} and \ref{uv-noreg} the
resulting brightness distribution from a fit to the B and UV band OM
data in a case where {\sl no} regularization was applied. The fit is
excellent, but the resulting brightness distribution suffers slightly
from the `scissors' effects, i.e. the point spread function of the map
has a scissors shape. This is because there is not enough information
in the data to uniquely locate point like emission. Thus point source
emission is spread out along the lines that mark the shadow of the
secondary at the ingress and egress phases.  This is a well known
effect and was demonstrated by Horne (1985).

Using the regularization described earlier we obtain the fits for B
and UV bands shown in Figures \ref{breg} and \ref{uvreg}
respectively.  The coordinate system for the fireflies is a Cartesian
one, where the primary star is located in the origin and the X-Y plane
is the orbital plane.  The coordinates of the L1 point are roughly
(0.7,0,0) and a ballistic freefall stream is overplotted. The observer
lies `above' the orbital plane, i.e. she/he would have a positive
Z-coordinate.

Figure \ref{fly-hist} shows the radial histograms of the
distribution of flies. The resulting brightness distributions seem to indicate that whilst
the UV emission comes mostly from the white dwarf or the boundary
layer and the inner disc, the optical B band emission also contains a
faint component from the accretion stream - disc interactions,
i.e. the hot spot (and maybe even from the freefall part of the
stream). This is also rather obvious directly from the light curve
shapes, given the difference in eclipse assymmetry between the two
passbands.  

Furthermore, we find that the hot spot is located exactly on the
ballistic trajectory, as expected. The distance is $\sim$ 0.3a from the white
dwarf, which confirms the values reported by Wood et al. (1989).  
In addition to this we find marginal evidence for
optical emission from the accretion stream prior to the main hot spot
impact region. Our results show, that in the B band 90\% of the
emission comes within a distance of 0.29a from the white dwarf and
95\% within 0.33a. This implies that the hot spot contributes roughly 
5\% of the total B band emission. The same radii in UV band are slightly 
greater in  comparison (0.32a and 0.38a). At face value this implies that the B
band radial emission distribution is concentrated more towards the
white dwarf than in UV passband.  This difference could be
due to the fact that whilst our UV data was taken 4 days after an
outburst, the B band data was taken more than a month later. Dwarf novae
discs are thought to be more uniformly bright during the outbursts.
Although the $V$ band mag was similar in both observations this does not
preclude a difference in the size of the accretion disc in the two
observations.

\section{Discussion}

\subsection{The origin of the X-ray modulation}
\label{mod_discuss}

Ramsay et al (2001a) found that in X-rays the duration of the eclipse
ingress and egress is 20--30sec, which is roughly the same time as
that expected for the secondary star to eclipse the diameter of the
white dwarf. These authors concluded that the X-ray emission region is
hence extended in white dwarf longitude, probably centered near the
spin equator. However, Wheatley \& West (2003), using a re-analysis of
the same data, conclude that the X-ray emission originated close to
the upper pole on the white dwarf. In the latter scenario, an
accretion column, or curtain, of material will rotate on the spin
period of the white dwarf.

This is consistent by our spectral modelling in \S \ref{spec} which
indicates that the most likely cause of these quasi-stable modulations
is a variation in the absorption column over this period. {\sl If} the
2193 sec period was the spin period of the white dwarf, how would that
manifest itself as a variation in the absorption column?  In the case
of intermediate polars, some systems show an energy dependent X-ray
variation at the spin period.  This has been seen in EX Hya (Rosen,
Mason \& Cordova 1988, Cropper et al 2002) and is thought to be due to
observing the emission site through an accretion curtain whose
absorption column varies as the white dwarf rotates.

A second possibility may be due to a variation in the distance from
the inner edge of the accretion disc to the white dwarf. Further, this
distance could vary in azimuth in the binary frame. Thus the white
dwarf-accretion disc interaction would lead to a variation at a beat
period and could account for why the modulation is only seen for
around 1/3 of the observation duration. This hypothesis requires that
the 2193 sec period is the spin period of the white dwarf ($\omega$),
whilst the 3510 sec modulation is the beat period between the spin and
the 5436 sec orbital period ($\Omega$). The beat period based on the
2193 s spin period would be ($\omega-\Omega$) 3676 sec, which is
6.6$\sigma$ distant from the best fit period. However, if the spin
period is actually 1.7$\sigma$ shorter than the best fit value (giving
2133 sec) then the observed 3510 sec period would be exactly the
expected $\omega-\Omega$ beat period. In this scenario, any very short
periods (for instance the 17.94 and 18.15 sec periods reported by
Marsh \& Horne, 1998) would have to be explained in other terms, for
instance white dwarf oscillations.

Finally, Marsh \& Horne (1998) suggested that one of the $\sim$18 sec
periods could represent $\omega$, or alternatively that one of the
periods was the Keplerian period and the other was the beat period
between this and the spin period (implying a spin period of $\sim$1550
sec). If $\sim$1550 sec was the spin period, then most of the
amplitude peaks in our Figure \ref{pow-signi} could be explained as
naturally occuring sidebands of the spin and orbital frequencies in
the following manner: 1248 sec ($\omega + \Omega$), 1574 sec
($\omega$), 2193 sec ($\omega-\Omega$), 4844 sec
(2($\omega-\Omega$)). The main drawback with this hypothesis is that
there is no simple explanation for the very significant 3510 sec
period. Also, it is possible that not all of the 5 periods mentioned
above are in fact physical (cf \S 3).

Recently, the short period variability in cataclysmic variables has been 
extensively discussed in a series of papers by Warner \& Woudt (most
recently in Warner, Woudt \& Pretorius (2003)). In conclusion they have
found that typically $P_{QPO} \sim 15 P_{DNO}$. In addition they also
find second type of DNO's with: $P_{lpDNO} \sim 4 P_{DNO}$.  Assuming 
that one of the periods seen here, 2193s or 3500s, would be a QPO
period the above mentioned relationship would then predict DNO periods
around 150s and 230s respectively. There is no evidence for such 
periods in our data. However, this is expected since the DNO and QPO
periods are normally found during outbursts, and during our observation
the source was in quiescence.   

We conclude that based on our data, we cannot provide a unique
explanation for the 2193 sec and 3510 sec periods. We can, however,
conclude that these periods occur due to changes in the line of sight
absorbing column.

\subsection{The Optical Light Curve Modelling}

Much previous work has been done on the eclipse profiles of OY Car
ranging from infrared to X-rays (see for instance Berriman 1984, Cook
1985, Krzeminski \& Vogt 1985, Schoembs 1986, Berriman 1987, Wood et
al. 1989, Rutten et al. 1992, Horne et al 1994, Bruch, Beele and
Baptista 1996, Pratt et al. 1999, Ramsay et al. 2001a, Wheatley \&
West 2003). Wood et al. (1989) use the optical eclipse profiles to
measure accurate system parameters and assuming a ballistic free fall
trajectory for the accretion stream they deduce a disc outer radius of
approximately 0.31a. Rutten et al. (1992) re-analysed the set of
eclipse profiles from Vogt (1983) that cover the rise into an
outburst. Their eclipse mapping analysis shows that whilst the
quiescence eclipse profiles are dominated by the white dwarf and the
hot spot, the rest of the disc becomes dominant in the outburst.  The
HST UV observations by Horne et al. (1994) demonstrate that the
eclipse profiles become more symmetric and `tophat shaped' towards the
shorter wavelengths. This indicates that the white dwarf and/or the
boundary layer is the major source of emission in the UV, whilst the
hot spot contribution rises towards the optical wavelengths.

Our light curve modelling confirms the previous results on the origin
of the quiescent eclipse profiles. In particular they indicate, that
both in B and UV bands, the majority of the emission arises close to
the white dwarf, possibly either the boundary layer or the white dwarf
itself.  This is also supported by the ingress and egress time
analysis of the the same B, UV dataset by Ramsay et al. (2001a).

The freefall accretion stream and especially the hot spot, where the
stream hits the disc outer edge is mainly emitting in the B band and
hardly visible in the UV maps.  Our modelling also suggests that in
addition to the hot spot, the ballistic accretion stream might
contribute to the B band emission as well.

Our novel approach allows simultaneous fitting of the eclipse profiles
and orbital light curves of cataclysmic variables. This removes bias
introduced by artificially having to remove the orbital modulation
prior to the eclipse profile mapping. Furthermore, the `firefly
approach' has potential to be easily converted into full 3D mapping,
if a proper regularisation can be employed.

\section{Conclusions}

We have reanalysed the {\xmm} observations of the dwarf novae OY
Car. In particular, we have reanalysed the X-ray light curve and
spectrum and developed a new model to fit the OM B band and UV data.

We have investigated the X-ray light curve over the time interval
where the previously reported 2240 sec modulation was strongly seen.
This has allowed us to refine this period to 2193 sec. We also detect
a modulation at a period of 3510 sec which we also believe to be
significant. We have performed phase resolved X-ray spectroscopy on
these periods and find that the soft X-ray light curve is
anti-correlated with the absorption column density. This implies that
the light curve modulation is due to us observing the X-ray emission
region through a rotating accretion column or curtain. Although, not
conclusive, this supports the hypothesis of Ramsay et al (2001a) that
the 2193 sec is the spin period of the white dwarf. Furthermore, this
is in agreement with findings of Wheatley \& West (2003), where they
conclude that the X-ray emission probably originates near the polar 
region on the white dwarf surface.

To model the OM B and UV light curves, we have applied the fireflies
model of Hakala et al (2002). We simulated light curves using a given
accretion disc brightness distribution and used our technique to fit
the synthetic light curves. The resulting distribution of fireflies is
encouraging. We then applied the model to the OM data and find that
that the hot-spot, where the accretion stream hits the disc, is
exactly on the ballistic trajectory. Further, we find some evidence
that the B band emission is more concentrated than the UV emission,
although this maybe due to the fact that the B band data were taken 1
month after the UV data. Our technique can be applied to other disc
accreting systems.

\begin{acknowledgements}
This paper is based on observations obtained with XMM-Newton, an ESA
science mission with instruments and contributions directly funded by
ESA Member States and the USA (NASA). PJH is an Academy of Finland
research fellow.
\end{acknowledgements}

\end{document}